\documentclass[a4paper,UKenglish,runningheads,11pt]{llncs}

\usepackage{tabularx,booktabs,multirow,delarray,array}
\usepackage{graphicx,amssymb,amsmath,amssymb,mathtools,amsfonts}
\usepackage{enumerate}
\usepackage[ruled,vlined,linesnumbered]{algorithm2e}
\usepackage{wrapfig}
\usepackage{latexsym}
\usepackage{lineno}
\usepackage{hyperref}
\usepackage{subcaption}
\usepackage{float}
\usepackage{url}
\usepackage{upgreek}
\usepackage{textgreek}
\usepackage{caption}
\usepackage{subcaption}
\usepackage{enumitem}
\usepackage{fullpage}

\newenvironment{proof}{\par\noindent{\bf Proof:}}{\mbox{}\hfill$\qed$\\}

\newlength{\bibitemsep}
\newlength{\bibparskip}
\let\oldthebibliography\thebibliography
\renewcommand\thebibliography[1]{
  \oldthebibliography{#1}
  \setlength{\parskip}{\bibitemsep}
  \setlength{\itemsep}{\bibparskip}
}

\newcommand{\ignore}[1]{ }

\newcounter{rem}
\def\etal{\textsl{et~al.}}

\def\qed{\hbox{\rlap{$\sqcap$}$\sqcup$}}

\begin{document}

\title{Two-point Approximate Shortest Path Queries among Convex Polygonal Obstacles in the Plane}
\titlerunning{Computing a two-point approximate shortest path}

\author{
Siddharth Gaur\inst{1}
\and
R. Inkulu\inst{1}
}

\institute{
Department of Computer Science and Engineering\\
Indian Institute of Technology Guwahati\\
\email{\{sgaur,rinkulu\}@iitg.ac.in}
}

\authorrunning{S. Gaur and R. Inkulu}

\maketitle

\pagenumbering{arabic}
\setcounter{page}{1}

\begin{abstract}
Given a polygonal domain $\cal P$ consisting $h$ pairwise disjoint convex polygonal obstacles together defined with $n$ vertices and a positive real number $\epsilon$ in $(0, 0.6)$, this paper presents an algorithm to preprocess $\cal P$ in $O(n+\frac{h}{\epsilon}(h+\frac{1}{\sqrt{\epsilon}})\lg(\frac{h}{\sqrt{\epsilon}}))$ time to compute data structures of size $O(n+\frac{h}{\sqrt{\epsilon}} (h+\frac{1}{\epsilon}))$ so that given any two points $s$ and $t$ in the free space defined by $\cal P$, a path between $s$ and $t$ with a $(1+\epsilon)$ multiplicative stretch and $13\ell$ additive stretch is output in $O(\frac{1}{\sqrt{\epsilon}}(\lg{\frac{h}{\sqrt{\epsilon}}})+\frac{h}{\epsilon^{2.5}}(\lg{\lg(\frac{h}{\sqrt{\epsilon}})}))$ time.
Here, $\ell$ is upper bounded by $(\sqrt{2\epsilon}) (\max_{P_i \in \cal P} \max_{p, q \in P_i} |pq|)$.
\end{abstract}

%\keywords{Computational Geometry; Shortest Paths; Approximation Algorithms.}

\section{Introduction}
\label{sect:intro}

A {\it polygonal domain} $\cal P$ in $\mathbb{R}^2$ is a finite collection of pairwise-disjoint simple polygonal obstacles. 
The free space $\cal{F(P)}$ of $\cal P$ is the plane that contains obstacles in $\cal P$, excluding the interiors of those obstacles.
Given a polygonal domain $\cal P$ comprising $h$ polygonal obstacles defined with $n$ vertices and two points $s$ and $t$ in the $\cal{F(P)}$, the Euclidean shortest path finding problem seeks to compute a shortest path between $s$ and $t$ that lies in $\cal{F(P)}$.
If every obstacle in $\cal P$ is convex, then $\cal P$ is called a {\it convex polygonal domain}.

Computing Euclidean shortest paths in polygonal domains is a well-studied problem in computational geometry.
There are primarily two approaches for efficiently computing shortest paths in polygonal domains.
The first approach involves constructing a visibility graph in $\cal{F(P)}$.
Any two vertices $v', v''$ of $\cal P$ are said to be mutually visible whenever the relative interior of line segment $v'v''$ is located in the $\cal{F(P)}$.
The nodes of the visibility graph of $\cal P$ comprise the vertices of $\cal P$, and there is an edge between any two nodes whenever those two nodes are mutually visible.
It is shown that a shortest path in $\cal{F(P)}$ between any two vertices of $\cal P$ is a shortest path in the visibility graph.
This reduces the problem of computing a geodesic shortest path in $\cal{F(P)}$ to a graph-theoretic problem.
The algorithms in \cite{journals/siamcomp/KapoorM00,journals/dcg/KapoorMM97,journals/ipl/Welzl85} use this approach.
Several other algorithms for computing visibility graphs and their characterizations are presented in the monograph by Ghosh~\cite{books/visalgo/skghosh2007}.
The second approach propagates the continuous Dijkstra wavefront in $\cal{F(P)}$ from the given source vertex to the destination vertex.
In this approach, a shortest path wavefront originating at the source expands through $\cal{F(P)}$ until the wavefront strikes the destination. 
Using this approach, Hershberger and Suri~\cite{journals/siamcomp/HershbergerS99} presented an $O(n \lg{n})$ time algorithm. 
Kapoor's earlier algorithm~\cite{conf/stoc/Kapoor99}, extended by Inkulu~\etal~\cite{journals/corr/InkuluKM10}, runs in $O(n + h((\lg h)^\delta + (\lg{n})(\lg h)))$ time, where $\delta$ is a small positive constant that arises from triangulating $\cal{F(P)}$ using the algorithm by Bar-Yahuda and Chazelle~\cite{journals/ijcga/Bar-YehudaC94}.
Numerous studies have addressed the shortest path computations in polygonal domains~\cite{journals/algorithmica/AsanoAGHI86,conf/soda/ChiangM99,journals/algorithmica/GuibasHLST87,journals/comgeo/InkuluK09a,conf/isaac/InkuluKap19,conf/socg/OverWelzl88,journals/siamcomp/SharirS86,journals/jacm/StorerR94}. 
In three dimensions, the problem is significantly more complex.
Canny and Reif~\cite{conf/focs/CannyR87} showed that computing shortest paths among polyhedral obstacles in $\mathbb{R}^3$ is NP-hard. 
A survey of algorithms for shortest paths in geometric domains is provided by Mitchell~\cite{coll/hb/Mitch17}.
The shortest path algorithms rely on many fundamental data structures and algorithms for geometric problems.
These are detailed in the standard textbooks for computational geometry, by Preparata and Shamos~\cite{books/compgeom/prep1985} and de~Berg \etal~\cite{books/compgeom/deberg2008}.

The following are three main variants of the Euclidean shortest path finding problem in polygonal domains: 
(i) both $s$ and $t$ are given as input with $\mathcal{P}$, 
(ii) only $s$ is provided as input with $\mathcal{P}$, and 
(iii) neither $s$ nor $t$ is given as input.
Type~(i) problems are single-shot and involve no preprocessing. 
The preprocessing phase of the algorithm for a type (ii) problem constructs a shortest path map with $s$ as the source so that a shortest path between $s$ and any given query point $t$ can be found efficiently.
In the third variation, known as the two-point shortest path query problem, the polygonal domain $\mathcal{P}$ is preprocessed to construct data structures that facilitate answering shortest path queries between any given pair of query points $s$ and $t$.
Naturally, the last variant is considered the hardest among these three.
In this, not only the preprocessing time but also the space required by the data structures computed during the preprocessing phase, and the time to answer shortest path queries are important.
In this paper, we devise an efficient approximation algorithm for this third variant.

The two-point shortest path query problem within a given simple polygon was addressed by Guibas and Hershberger~\cite{journals/jcss/GuibasH89}.
It preprocessed the simple polygon in $O(n)$ time and constructed data structures of size $O(n)$ to answer two-point shortest distance queries in $O(\lg{n})$ time.
Exact two-point shortest path queries in the polygonal domain were explored by Chiang and Mitchell~\cite{conf/soda/ChiangM99}.
One of the algorithms in \cite{conf/soda/ChiangM99} constructs data structures of size $O(n^5)$ and answers two-point shortest path distance queries in $O(h + \lg{n})$ worst-case time.
And, another algorithm in \cite{conf/soda/ChiangM99} builds data structures of size $O(n+h^5)$ and outputs any two-point distance query in $O(h\lg{n})$ time.
Guo~\etal~\cite{conf/aaim/GuoMS08} preprocessed $\cal{F(P)}$ in $O(n^2\lg{n})$ time to compute data structures of size $O(n^2)$ for answering two-point distance queries for any given pair of query points in $O(h\lg{n})$ time.
In all of these algorithms, an approximate shortest path itself is found in additional time $O(k)$, where $k$ is the number of edges in the output path.

Given the difficulty of answering two-point shortest path queries in polygonal domains, various approximation algorithms have been devised to output an approximate shortest path between the two query points instead of an optimal one.
For any two points $p$ and $q$ in the free space of a polygonal domain, let $d(p, q)$ be the shortest path distance between $p$ and $q$.
Then, if the length of the path output by the algorithm is at most $\alpha \cdot d(p, q) + \beta$, then $\alpha$ is called the {\it multiplicative stretch} and $\beta$ is called the {\it additive stretch} of the path produced.
When there is no additive stretch, the {\it (multiplicative) stretch} of a path output by an algorithm is the maximum, over all pairs of points $p$ and $q$, the ratio between the length of the path between $p$ and $q$ output by the algorithm to the shortest path distance between $p$ and $q$.
Clarkson first made such an attempt in \cite{conf/stoc/Clarkson87}.
Chen~\cite{conf/soda/Chen95} used the techniques from \cite{conf/stoc/Clarkson87} in constructing data structures of size $O(n\lg{n} + \frac{n}{\epsilon})$ in $o(n^{3/2})+O(\frac{n}{\epsilon}\lg{n})$ time to output $(6+\epsilon)$-approximate two-point distance queries in $O(\frac{1}{\epsilon}\lg{n}+\frac{1}{\epsilon^2})$ time.
Arikati~\etal~\cite{conf/esa/ArikatiCCDSZ96} devised a family of algorithms to answer two-point approximate distance queries.
Their first algorithm outputs a $(\sqrt{2}+\epsilon)$-approximate distance; depending on a parameter $1 \le r \le n$, in the worst-case, either the preprocessed data structures of this algorithm take $O(n^2)$ space or the query time is $O(\sqrt{n})$. 
Their second algorithm takes $O(n)$ time per query to report an $(\sqrt{2}+\epsilon)$-approximation of the shortest path distance.
The stretch of the third and fourth algorithms proposed in \cite{conf/esa/ArikatiCCDSZ96} are respectively  $(2\sqrt{2}+\epsilon)$ and $(3\sqrt{2}+\epsilon)$.
Again, in all of these algorithms, a shortest path itself is found in additional time $O(k)$, where $k$ is the number of edges in the output path.
When $p$ and $q$ are given with $\cal P$, Agarwal~\etal~\cite{conf/soda/AgarwalSY09} computes a $(1+\epsilon)$-approximate geodesic shortest path between $p$ and $q$ in $O(n + \frac{h}{\sqrt{\epsilon}}\lg(\frac{h}{\epsilon}))$ time when the obstacles are convex.

Another related problem is the computation of geometric spanner networks.
Given a graph $G$, a subgraph $H$ of $G$ is called a $t$-spanner for $t \geq 1$ if for every pair of vertices $u$ and $w$, the shortest path between $u$ and $w$ in $H$ is at most $t$ times the length of the shortest path between $u$ and $w$ in $G$.
When the graph is embedded in a geometric domain, such a spanner is called a geometric spanner network. 
The spanner networks for point sets in the free space of a polygonal domain were studied in Clarkson~\cite{conf/socg/ClarksonKV87}, Chen~\cite{conf/soda/Chen95}, and Arikati~\etal~\cite{conf/esa/ArikatiCCDSZ96}. 
Yao-graphs~\cite{journals/siamcomp/Yao82} and \textTheta-graphs~\cite{conf/stoc/Clarkson87} are commonly used in computing geometric spanner networks and these are used in computing more sophisticated spanner networks. 
Narasimhan and Smid~\cite{books/compgeom/narsmid2007} and Bose and Smid~\cite{journals/cgta/BoseSmid13} provide detailed surveys on geometric spanner networks.

\subsection{Our Contributions}

Our preprocessing algorithm first computes a convex polygonal domain $\cal Q$ from the input convex polygonal domain $\cal P$ so that $\cal Q$ is defined with $O(\frac{h}{\sqrt{\epsilon}})$ vertices.
Here, $\epsilon$ is the input parameter and $h$ is the number of obstacles in $\cal P$.
This involves computing a convex polygon $Q_i$ corresponding to each convex polygon $P_i$ in $\cal P$ such that $Q_i$ contains $P_i$ and the line containing any edge of $Q_i$ is a tangent to $P_i$.
The convex polygonal domain ${\cal Q} = \bigcup_i Q_i$ is obtained by simplifying each of the obstacles in $\cal P$. 
Specifically, $\cal Q$ is computed using an algorithm from Agarwal~\etal~\cite{conf/soda/AgarwalSY09} so that $\cal Q$ ensures $d_{\cal Q}(s, t) \leq (1+\epsilon) d_{\cal P}(s, t)$ for any two points $s, t \in \cal{F}(Q)$.
Here, $d_{\cal Q}(s, t)$ (resp., $d_{\cal P}(s, t)$) is the shortest distance between $s$ and $t$ in $\cal{F(Q)}$ (resp., $\cal{F(P)}$).
The efficiency in computing a routing path from $s$ to $t$ is achieved due to the lesser complexity of $\cal Q$, in comparison to the complexity of $\cal P$.
Unlike in Inkulu and Kumar~\cite{journals/ijfcs/InkuluKum24}, where ${\cal{F(P)}} \subseteq {\cal{F(Q)}}$, the sketch $\cal Q$ of $\cal P$ used here enforces ${\cal{F(Q)}} \subseteq {\cal{F(P)}}$.

Based on $\cal Q$, we partition the boundary of each obstacle in $\cal P$ into contiguous sections, called patches.
For every $Q_i \in \cal Q$, for each vertex $v'$ of $Q_i$, vertices on the $bd(P_i)$ that are visible to $v'$ are contiguous on the $bd(P_i)$, and the section of boundary due to these vertices is called a patch of $P_i$ associated to $v'$. 
And the simple polygon in $\cal{F(P)} \backslash \cal{F(Q)}$ bounded by two edges of $Q_i$ incident to $v'$ and the patch associated to $v'$ is called a pocket of $v'$.
Noting every point in the pocket of $v'$ is visible to $v'$, to compute an approximate shortest path between two points $s$ and $t$, wherein $s$ is located in the pocket of $v'$, our algorithm outputs the line segment $sv'$ concatenated with an approximate shortest path from $v'$ to $t$. 
Analogously, if $t$ is located in the pocket of $w'$, the line segment incident to $t$ in the approximate shortest path output by our algorithm is $w't$. 
Thus, we reduce the problem of computing an approximate shortest path (between $s$ and $t$) in $\cal{F(P)}$ to the problem of computing an approximate shortest path (between $v'$ and $w'$) in $\cal{F(Q)}$. 

Next, we detail the other structures computed in the preprocessing phase, which together facilitate the computation of an approximate shortest path between any two vertices of $\cal Q$.
Following Clarkson~\cite{conf/stoc/Clarkson87}, we introduce a set $\cal C$ of $\frac{1}{\epsilon}$ cones with apex at the origin of the coordinate system, each with cone angle $\epsilon$ except for one with cone angle $2\pi-\lfloor \frac{2\pi}{\epsilon} \rfloor \epsilon$.
By using these cones, our algorithm constructs $|{\cal C}|$ number of conical Voronoi diagrams (CVDs), each corresponding to a cone in $\cal C$.
As in Inkulu and Kapoor~\cite{conf/isaac/InkuluKap19}, noting every maximal line segment in a shortest path in $\cal{F(Q)}$ between any two points that are not mutually visible is tangent to a convex polygon, we further reduce the total number of cones introduced per obstacle to $\frac{1}{\epsilon}$.
Further, following Banyassady~\etal~\cite{journals/cgta/BanyaCKMR20}, with respect to each vertex $v'$ of $\cal Q$ and a cone $C_{v'}$ in $\cal C$ translated so that its apex is at $v'$, we partition the $bd({\cal Q})$ into contiguous sections called intervals, wherein the successor along a shortest path from $v'$ to every vertex $v''$ that belongs to an interval $I$ is the same vertex of $\cal{F(Q)}$ and $v'v''$ belongs to $C_{v'}$.
We associate with $v'$, the first and last vertices of interval $I$ together with $u'$, where $u'$ is the closest vertex visible to $v'$ in $C_{v'}$.
These intervals help in efficiently determining the subsequent line segment on the approximate shortest path in $\cal{F(Q)}$ based on the interval to which a destination vertex belongs. 
However, unlike in \cite{journals/cgta/BanyaCKMR20}, we define these intervals on the $bd({\cal Q})$, in place of defining them on the $bd({\cal P})$.
To determine whether a given query point $s$ (resp., $t$) is located in $\cal{F(Q)}$ or to identify a specific pocket to which it belongs in $\cal{F(P)} \backslash \cal{F(Q)}$, we augment Kirkpatrick's point location data structure \cite{journals/siamcomp/Kirkpatrick83} for a triangulation of $\cal Q$ with one triangle per pocket.
By following Chen~\cite{conf/soda/Chen95}, we compute a set of trapezoidal decompositions of $\cal{F(Q)}$ to facilitate determining whether the two query points in $\cal{F(Q)}$ are mutually visible.

In computing an approximate shortest path between $s'$ and $t'$ belonging to $\cal{F(Q)}$ in the query phase, for every cone $C^j$ in $\cal C$, using the CVD of $C^j$, we find a closest vertex $v_s^j$ (resp., $v_t^j$) of $bd({\cal Q})$ that is visible to $s$ (resp., $t$) in $C_s$ (resp., $C_t$).
Let $Q_{s'}$ (resp., $Q_{t'}$) be the set of closest visible vertices to $s'$ (resp., $t'$) wherein each point in $Q_{s'}$ (resp., $Q_{t'}$) correspond to a distinct cone in $\cal C$.
For every pair of points in $Q_{s'} \times Q_{t'}$, we compute an approximate shortest path using data structures computed in the preprocessing phase and output a path with the shortest distance.

The efficiency of our algorithm is due to patches defined on the $bd({\cal P})$, a collection of pockets defined by each of these patches, intervals computed on the $bd({\cal Q})$ with respect to each vertex of $\cal Q$, conical Voronoi diagrams to help in finding vertices of $\cal Q$ that are closest to any given query point in any cone in $\cal C$, augmented triangles to facilitate point location, and the interval specific data structures associated to any vertex $v$ of $\cal Q$ to store the subsequent line segment along an approximate shortest path from $v$.

\begin{theorem}
Given a convex polygonal domain $\cal P$ comprising $h$ convex polygonal obstacles defined with $n$ vertices and a real number $\epsilon \in (0, 0.6)$, in $O(n+\frac{h}{\epsilon}(h+\frac{1}{\sqrt{\epsilon}})\lg(\frac{h}{\sqrt{\epsilon}}))$ time, the preprocessing algorithm computes data structures of size $O(n+\frac{h}{\sqrt{\epsilon}} (h+\frac{1}{\epsilon}))$, so that to output in time $O(\frac{1}{\sqrt{\epsilon}}(\lg{\frac{h}{\sqrt{\epsilon}}})+\frac{h}{\epsilon^{2.5}}(\lg{\lg(\frac{h}{\sqrt{\epsilon}})}))$ an approximate shortest path between any two query points in $\cal{F(P)}$ wherein that path has multiplicative stretch $1+\epsilon$ and an additive stretch $13 \ell$.
Here, $\ell$ is upper bounded by $(\sqrt{2\epsilon})(\max_{P_i \in {\cal P}} \max_{p, q \in P_i} |pq|)$.
\end{theorem}

\begin{table}[!ht]
\begin{tabularx}{\textwidth} {  
  | >{\raggedright\arraybackslash}p{2.55cm}
  | >{\raggedright\arraybackslash}p{2.5cm}
  | >{\raggedright\arraybackslash}p{3.2cm}
  | >{\raggedright\arraybackslash}p{2.75cm}
  | >{\raggedright\arraybackslash}X
  | }
  \hline

\hline
\textbf{} 
& Stretch
& Preprocessing time
& Space
& Query time \\
\hline
Chiang and Mitchell~\cite{conf/soda/ChiangM99}
    & optimal 
    & $-$
    & $O(n^{11})$ 
    & $O(k+\lg{n})$ \\

\quad 
    & optimal 
    & $-$
    & $O(n^{10}\lg n)$ 
    & $O(k+(\lg n)^2)$ \\

\quad 
    & optimal 
    & $-$
    & $O(n^{5})$ 
    & $O(k+h+(\lg{n}))$ \\

\quad 
    & optimal 
    & $-$
    & $O(n + h^5)$ 
    & $O(k+h\lg{n})$ \\

\hline

Guo~\etal~\cite{conf/aaim/GuoMS08}
    & optimal 
    & $O(n^2 \lg n)$ 
    & $O(n^2)$ 
    & $O(k+h\lg n)$ \\

\hline

Chen~\etal~\cite{journals/ijcga/ChenDK01}
   & optimal 
   & $O(n^2)$ 
   & $O(n^2)$ 
   & $O(k+\min(|Q_s|,|Q_t|)\lg n)$ \\

\hline

Chen~\cite{conf/soda/Chen95}
    & $6+\epsilon$ 
    & $O(\frac{n\lg n}{\epsilon} + \frac{q^{3/2}}{\sqrt{\lg q}})$ 
    & $O(n\lg n + \frac{n}{\epsilon})$ 
    & $O(k+\frac{\lg n}{\epsilon} + \frac{1}{\epsilon^2})$ \\

\hline

Arikati~\etal~\cite{conf/esa/ArikatiCCDSZ96}
    & $\sqrt{2}+\epsilon$ 
    & $O(n^2/\sqrt{r})$ 
    & $O(n^2/\sqrt{r})$ 
    & $O(k+(\lg{n}) + \sqrt{r})$ \\

    & $\sqrt{2}+\epsilon$ 
    & $O(n\lg n)$ 
    & $O(n)$ 
    & $O(n)$ \\

    & $2\sqrt{2}+\epsilon$ 
    & $O(n^{3/2})$ 
    & $O(n^{3/2})$ 
    & $O(k+\lg{n})$ \\

    & $3\sqrt{2}+\epsilon$ 
    & $O(n^{3/2}+(\lg n)^{1/2})$ 
    & $O(n\lg{n})$ 
    & $O(k+\lg{n})$ \\

\hline

This result
    & $(1+ \epsilon)d(s,t) +$ $(13 \ell)$
    & $O(n+\frac{h}{\epsilon}(h+\frac{1}{\sqrt{\epsilon}})\lg(\frac{h}{\sqrt{\epsilon}}))$
    & $O(n+\frac{h}{\sqrt{\epsilon}} (h+\frac{1}{\epsilon}))$
    & $O(\frac{1}{\sqrt{\epsilon}}(\lg{\frac{h}{\sqrt{\epsilon}}})+\frac{h}{\epsilon^{2.5}}(\lg{\lg(\frac{h}{\sqrt{\epsilon}})}))$ \\

\hline
\end{tabularx}

\caption{\footnotesize Illustrates preprocessing time, space of preprocessed data structures, and the query times for two-point (approximate) shortest path queries.
Here, $k$ is the number of line segments in the output path, 
$r$ is an arbitrary integer such that $1 \le r \le n$, and 
$Q_s$ (resp., $Q_t$) is the size of the visibility polygon of $s$ (resp., $t$) with $O(1) \le \min(|Q_s|, |Q_t|) \le O(n)$.
(The visibility polygon of a point $q$ in $\cal{F(P)}$ is the simple polygon comprising all the points in $\cal{F(P)}$ that are visible to $q$.)
}
\label{table:compar}
\end{table}

As mentioned, four algorithms proposed by Chiang and Mitchell~\cite{conf/soda/ChiangM99} compute optimal shortest paths, but their space complexities are high.
Refer to Table~\ref{table:compar}.
On the other hand, algorithms proposed in both Guo~\etal~\cite{conf/aaim/GuoMS08} and Chen~\cite{journals/ijcga/ChenDK01} output an optimal two-point shortest path, but the space of the data structures computed is quadratic in $n$. Their respective query times are $O(h\lg{n})$ and $O(n\lg{n})$.
The preprocessing time and space complexities of Chen~\cite{conf/soda/Chen95} are super-linear in $n$, whereas the stretch of the path output is $6+\epsilon$.
Arikati~\etal~\cite{conf/esa/ArikatiCCDSZ96} proposed four approximation algorithms, but their preprocessing times are again super-linear in $n$; for three of these algorithms, the size of data structures is also $\omega(n)$. 
The stretch factors achieved in \cite{conf/esa/ArikatiCCDSZ96} do not have additive factors, but the multiplicative stretch is larger than $1+\epsilon$.
The algorithm proposed herewith has preprocessing time and space that are linear in $n$ and quadratic in $h$.
And the query time of this result has $h$ multiplied by a term $o(\frac{1}{\epsilon^{2.5}}\lg{n})$. 
However, our algorithm also has an additive approximation factor.
And the algorithm proposed herewith is for convex polygonal domains, and all other results mentioned in Table~\ref{table:compar} are for polygonal domains whose obstacles are not necessarily convex.

\section{Preliminaries}
\label{subsect:prelim}

We denote the input convex polygonal domain by $\cal P$.
The $\cal P$ is assumed to consist of $h$ convex polygonal obstacles, together defined with $n$ vertices.
We assume obstacles in $\cal P$ are in general position, that is, no two vertices of $\cal P$ have the same $x$- or $y$-coordinates, and no three vertices are collinear.
As mentioned, the free space $\cal{F(P)}$ of $\cal P$ is the plane containing obstacles in $\cal P$, excluding the interiors of those obstacles.
For every $1 \le i \le h$, the $i^{th}$ obstacle is denoted by $P_i$, and its boundary is denoted by $bd(P_i)$.
The boundary of $\cal P$ is the $\bigcup_i bd(P_i)$.
We denote the number of vertices of $P_i$ by $n_i$.
%Every vertex $v_{ik}$ of $\cal P$ is associated with two positive integers, $i$ to denote the obstacle to which it belongs and $k \in [1, n_i]$ to uniquely identify the vertex on the $bd(P_i)$.

Two points $p$ and $q$ in $\cal{F(P)}$ are said to be {\it visible} if the interior of the line segment $pq$ does not intersect the $bd(P_i)$ for any $i$.
The length of a polygonal path $\pi$ in $\cal{F(P)}$ is the sum of lengths (measured in Euclidean metric) of all the line segments belonging to $\pi$.
A {\it geodesic path} in $\cal{F(P)}$ is a simple polygonal path in which every pair of successive vertices is mutually visible, and whose length cannot be further reduced by small perturbations.
Among all such paths between two points $p$ and $q$, a path of minimum length is called a {\it shortest (geodesic) path}, denoted by $\pi_{\cal P}(p,q)$.
The length of $\pi_{\cal P}(p,q)$ is denoted by $d_{\cal P}(p, q)$. 
When the context is clear, we denote $\pi_{\cal P}(p, q)$ by $\pi(p, q)$ and $d_{\cal P}(p, q)$ by $d(p, q)$. 
Though there may be more than one shortest path between any two points, for the convenience of presenting the algorithm, we assume that there is a unique shortest path.
The Euclidean distance between any two points $p$ and $q$ in $\mathbb{R}^2$ is denoted by $\vert pq \vert$.
%The coordinates of any point $p$ in $\mathbb{R}^2$ is denoted by $coord(p)$.

Let $r'$ and $r''$ be two non-parallel rays with origin at a point $p$.
Let $\overrightarrow{v_1}$ and $\overrightarrow{v_2}$ be the unit vectors along rays $r'$ and $r''$, respectively.
A {\it cone} $C_p(r', r'')$ is the set of points defined by rays $r'$ and $r''$ such that a point $q \in C_p(r', r'')$ if and only if $q$ can be expressed as a convex combination of vectors $\overrightarrow{v_1}$ and $\overrightarrow{v_2}$ with positive coefficients.
When the rays are evident from the context, we denote $C_p(r', r'')$ by $C_p$ or $C$.
The counterclockwise angle from the positive $x$-axis to the line that bisects $C_p$ is called the \emph{orientation} of $C_p$. 
The angle between rays $r'$ and $r''$ is the {\it cone angle} of $C_p$.
We now specialize this definition of a cone with respect to a vertex of a convex obstacle. 
Let $u, v, w$ be successive vertices of $\cal P$ that occur while traversing the boundary of $\cal P$.
Let $\alpha$ be the angle subtended by rays ${\overrightarrow uv}$ and ${\overrightarrow vw}$.
For each $j \in [0, k]$, we define $r_j(v)$ to be the ray obtained by rotating ${\overrightarrow vw}$ clockwise by an angle of $j \cdot \frac{\alpha}{k}$.
The cone $C_v(r_{j-1}(v), r_j(v))$ has vertex $v$ as its apex with $r_{j-1}(v)$ and $r_j(v)$ being the rays bounding it.
For the sake of simplicity, we denote $C_v(r_{j-1}(v),r_j(v))$ as $C_v^j$.
And we denote the set of all such cones at any vertex $v$ by $C(v)$, which is $\{C_v^j \mid 1 \leq j \leq k \}$.
We partition $\mathbb{R}^2$ by introducing a set $\cal C$ comprising $\frac{2\pi}{\epsilon}$ cones with each of their apex at the origin of the coordinate system and each of their cone angle being $\epsilon$ except one has cone angle $2\pi-\lfloor \frac{2\pi}{\epsilon} \rfloor \epsilon$.
For any cone $C$ in $\cal C$, with its apex translated to a point $p$ in $\mathbb{R}^2$ is denoted by $C_p$.

\section{Preprocessing Algorithm}

We first compute a sketch $\cal Q$ of $\cal P$. 
The $\cal Q$ is also a convex polygonal domain; however, $\cal Q$ simplifies $\cal P$ in the sense that the complexity of the former depends on only $h$ and $\epsilon$, and it is independent of $n$.
Based on $\cal Q$, we partition the $bd({\cal P})$ into contiguous sections, and for each section $\eta$, we make a vertex $v$ of $\cal Q$ the in-charge.
That is, for any vertex $u$ located on $\eta$, an approximate shortest path from any point to $u$ output by this algorithm passes through $v$.
Further, corresponding to $\eta$, we define a pocket in $\cal{F(P)} \backslash \cal{F(Q)}$ such that an approximate shortest path to any point in that pocket passes through $v$.
This reduces the problem of computing an approximate shortest path between two points in $\cal{F(P)}$ to the problem of computing an approximate shortest path in $\cal{F(Q)}$ between two vertices of $\cal Q$.
Subsection~\ref{subsect:sketch} details this part of the algorithm.

To efficiently compute an approximate shortest path between any two vertices of $\cal Q$ lying in the $\cal{F(Q)}$, we associate additional data structures to each vertex of $\cal Q$.
At any vertex $v'$ of $\cal Q$ that is incident on an approximate shortest path $\pi(s, t)$, the data structure associated to $v'$ help in efficiently finding the vertex $v''$ of $\cal Q$ such that $v''$ occurs after $v'$ along $\pi(s, t)$ such that the line segment $v'v''$ is in the $\cal{F(Q)}$.
Thus, we compute an approximate shortest path $\pi(s, t)$ incrementally by finding the next vertex at every vertex of $\cal Q$ incident to $\pi(s, t)$.
These are accomplished with conical Voronoi diagrams defined for $\cal{F(Q)}$ and the intervals defined on the $bd({\cal Q})$.
We enhance point location data structures for the triangulation of $\cal{F(Q)}$ to identify whether the given query points $s$ and $t$ are located in $\cal{F(Q)}$ or in $\cal{F(P)} \backslash \cal{F(Q)}$.
We also compute data structures to efficiently finding whether $s$ and $t$ located in $\cal{F(Q)}$ are mutually visible.
These additional data structures are detailed in Subsection~\ref{subsect:datastr}.

\subsection{Computing Pockets in $\cal{F(P)} \backslash \cal{F(Q)}$}
\label{subsect:sketch}

Given $\cal P$, the {\it sketch} $\cal Q$ of $\cal P$ is a convex polygonal domain such that $\cal Q$ simplifies the polygons in $\cal P$ so that the complexity of $\cal Q$ is a function of $h$ and $\epsilon$, instead of $n$.
There are two algorithms to compute such a sketch: one is in Agarwal~\etal~\cite{conf/soda/AgarwalSY09} and the other one was presented in Inkulu and Kapoor~\cite{conf/isaac/InkuluKap19}. 
For the former $\cal{F(Q)} \subseteq \cal{F(P)}$ whereas for the latter $\cal{F(Q)} \supseteq \cal{F(P)}$. 
Indeed, the latter was used in designing a routing algorithm in Inkulu and Kumar~\cite{journals/ijfcs/InkuluKum24}. 
Here we use the algorithm in \cite{conf/soda/AgarwalSY09}.

The algorithm in \cite{conf/soda/AgarwalSY09} computes a convex polygonal domain $\cal Q$ such that for every polygon $P_i$ in $\cal P$ there is a polygon $Q_i$ in $\cal Q$ such that 
(i) $Q_i$ contains $P_i$, 
(ii) every edge $e'$ of $Q_i$ passes through a vertex of $P_i$ so that a half-space induced by the line containing $e'$ has $P_i$,
(iii) for any two successive edges $e', e''$ along the $bd(Q_i)$, the angle subtended by lines orthogonal to $e'$ and $e''$ is upper bounded by $\sqrt{2\epsilon}$,
(iv) for any polygon $P_j$ in $\cal P$ with $P_j \ne P_i$, $Q_i$ does not intersect $P_j$, and 
(v) polygons in $\cal Q$ are pairwise disjoint.
They define a uniform sample $\cal N$ of $r$ directions, for a fixed constant $r$.
For every polygon $P_i \in \cal P$ and for every direction $u$ in $\cal N$, a line $u^\perp$ orthogonal to $u$ passing through a vertex of $P_i$ such that a half-plane defined by $u^\perp$ contains $P_i$ is defined.
Let $S$ be the set comprising all such half-planes.
For every two polygons $P_i$ and $P_j$ in $\cal P$ with a point $p_i \in bd(P_i)$ and a point $p_j \in bd(P_j)$ with line segment $p_ip_j$ not having any intersection with the interior of any polygons in $\cal P$, their algorithm introduces a line $\ell_{ij}$ such that one half-plane of $\ell_{ij}$ contains $P_i$ and the other one contains $P_j$.
An half-plane containing $P_i$ defined by a line $\ell_{ij}'$ parallel to $\ell_{ij}$ with a point of $bd(P_i)$ incident to $\ell_{ij}'$ is also included into $S$.
The intersection of all the half-planes in $S$ together defines $Q_i$. 
Since each such half-plane contains $P_i$, $Q_i$ also contains $P_i$.
Significantly, ${\cal F(Q)} \subseteq {\cal F(P)}$.
The polygons in $\cal Q$ are computed with a plane sweep algorithm.
The following lemma from \cite{conf/soda/AgarwalSY09} upper bounds the complexity of $\cal Q$.
And for any two points $s, t \in \cal{F(Q)}$, it relates $d_{\cal P}(s, t)$ and $d_{\cal Q}(s, t)$.

\begin{lemma}
\label{lem:agar}
Given $\cal P$ with $h$ convex polygonal obstacles defined with $n$ vertices, the sketch $\cal Q$ of $\cal P$ with $O(\frac{h}{\sqrt{\varepsilon}})$ number of vertices can be computed in $O(n+h\lg{h})$ time.
And, for any two points $s, t \in \cal{F}(Q)$, $d_{\cal P}(s, t) \le d_{\cal Q}(s, t) \le (1+\varepsilon) d_{\cal P}(s, t)$.
\end{lemma}

\begin{figure}[htbp]
  \centering
  \includegraphics[width=0.23\textwidth]{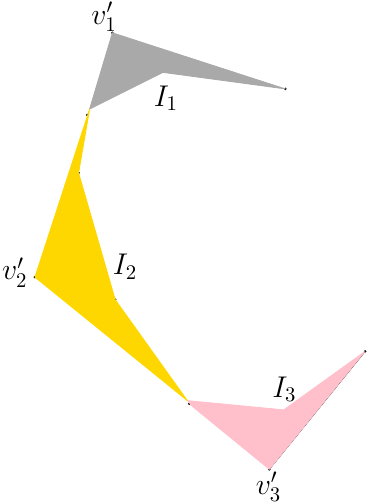}
  \caption{\footnotesize
  Illustrating three pockets located in $\cal{F(P)} \backslash \cal{F(Q)}$, each in a different colour.
  These pockets shown belong to an obstacle $Q_i$ of $\cal Q$, and the intervals corresponding to them are on the $bd(P_i)$ of $\cal P$, where $Q_i$ contains $P_i$.
  The representative vertex of interval $I_1$ (resp., $I_2, I_3$) comprising vertices on the $bd({\cal P})$ is vertex $v_1'$ (resp., $v_2', v_3'$) of $\cal Q$.
  \normalsize}
  \label{fig:pocket}
\end{figure}

For each obstacle $P_i \in \cal P$, the $bd(P_i)$ is partitioned into a collection of patches.
A patch of $P_i$ is a contiguous section of the $bd(P_i)$.
For each vertex $v'$ of $Q_i$ not in $P_i$, let $v_{ij}$ (resp., $v_{ik}$) be the vertex of $P_i$ that occurs first on the counterclockwise (resp., clockwise) traversal of $bd(Q_i)$ starting from $v'$.
The {\it patch} $\eta$ is the section of the $bd(P_i)$ that occurs in traversing the $bd(P_i)$ in the clockwise direction from $v_{ij}$ to $v_{ik}$, including $v_{ij}$ but not including $v_{ik}$.
We call $v' \in Q_i$ the {\it representative vertex} of $\eta$, and it is denoted by $rep(\eta)$.
And we call $v'$ the representative vertex of every vertex belonging to $\eta$.
We note that for every vertex $v$ of $P_i$ belonging to $\eta$, $v$ and $v'$ are mutually visible.
Refer to Fig.~\ref{fig:pocket}.
For every $Q_i \in \cal Q$ and $P_i \in \cal P$ with $P_i \subseteq Q_i$, $Q_i \backslash P_i$ is a collection of simple polygons, each of which has a patch on its boundary.
We call each such simple polygon a {\it pocket} of $Q_i$.
Each such pocket $P_{v'}$ has one vertex $v'$ that belongs to $Q_i$ but is not on the $bd(P_i)$.
We observe that $P_{v'}$ is a simple polygon bounded by the patch to which $v'$ is the representative and the two edges of $Q_i$ that are incident to $v'$.
We reduce the problem of finding an approximate shortest path in $\cal{F(P)}$ to the problem of finding an approximate shortest path in $\cal{F(Q)}$.
As part of this, for any point $s$ located in pocket $P_{v'}$, noting $s$ is visible to $v'$, the approximate shortest path between $s$ and any point located in $\cal{F(P)}$ output by this algorithm has the line segment $sv'$.
Since partitioning the $bd(P_i)$ into patches involve traversing the $bd(P_i)$ together with $bd(Q_i)$.
With these traversals, we could also identify pockets in $Q_i \backslash P_i$.
Considering $\cal P$ has $n$ vertices and $\cal Q$ has $O(\frac{h}{\sqrt{\epsilon}})$ vertices, the following lemma upper bounds the time involved in finding all the patches on the $bd({\cal P})$ and the pockets induced by them.

\begin{lemma}
\label{lem:pocketstime}
For every $i$, partitioning the boundary of every obstacle $P_i$ in $\cal P$ into patches and identifying all the pockets in $Q_i \backslash P_i$, together takes $O(n + \frac{h}{\sqrt{\epsilon}})$ time.
\end{lemma}

\subsection{Data Structures}
\label{subsect:datastr}

First, we describe {\it conical Voronoi diagrams} from Clarkson~\etal~\cite{conf/socg/ClarksonKV87}.
Let $C \in \cal C$ be a cone with orientation $\theta$, and let $C' \in \cal C$ be the cone with orientation $-\theta$.
For each cone $C \in \cal C$ and the set $K$ comprising vertices of $\cal Q$, the set of cones resulting from introducing a cone $C_p$ with its apex at point $p$ for every $p \in K$ is the conical Voronoi diagram $CVD(C, K)$ with respect to $C$ and $K$.
The $CVD(C, K)$ can be computed using a plane sweep algorithm (refer to de Berg~\etal~\cite{books/compgeom/deberg2008}) in $O(|K|\lg{|K|})$ time.
We associate a planar point location data structure with $CVD(C, K)$, enabling the location of any query point in the partition induced by $CVD(C, K)$.
Since $|{\cal C}|$ is $O(\frac{1}{\epsilon})$, the total time to compute $O(\frac{1}{\epsilon})$ CVDs and their corresponding point location data structures together is $O(\frac{1}{\epsilon} \frac{h}{\sqrt{\epsilon}} \lg(\frac{h}{\sqrt{\epsilon}}))$.
The total size of all CVDs is $O(\frac{1}{\epsilon} \frac{h}{\sqrt{\epsilon}})$.
If a query point $q$ belongs to $C_{v'}$ in $CVD(C, K)$ for a vertex $v'$ of $\cal Q$, then $v'$ is a {\it closest visible vertex} of $\cal Q$ to $q$ in $-C_q$.
Given a query point $q$ and a cone $C \in \cal C$, by using the planar point location data structure associated to $CVD(C, K)$, finding a closest visible vertex of $\cal Q$ to $q$ in $-C_q$ is reduced to finding a cone in $CVD(C, K)$ to which $q$ belongs. 
The later can be found in $O(\lg(\frac{h}{\sqrt{\epsilon}}))$ time.
If more than one point is closest in $C_v$ to $q$ (that is, $q$ is lying on an edge of $CVD(C, K)$), then we arbitrarily choose one of those vertices as the closest point in $C_v$ to $v$.

Let $v$ be any vertex of $\cal Q$.
Let $v', v, v''$ be the vertices that respectively occur while traversing the boundary of $P_i$ in counterclockwise order.
Also, let $C'$ be the cone defined by rays $\overrightarrow{vv'}, -\overrightarrow{vv''}$, and let $C''$ be the cone defined by rays $\overrightarrow{vv''}, -\overrightarrow{vv'}$.
We say any cone $C \in \mathcal{C}$ is {\it admissible} at $v$ whenever $C_v \cap C'$ or $C_v \cap C''$ is non-empty. 
Let $p$ and $q$ be two points in $\cal{F(Q)}$ such that $p$ and $q$ are not visible to each other, and a shortest path between $p$ and $q$ passes through $v$.
Since any shortest path is convex at $v$ with respect to $Q_i$, there is a shortest path between $p$ and $q$ such that one of its line segments lies in $C'$ and another line segment of that path lies in $C''$. 
Hence, as observed in Inkulu and Kapoor~\cite{conf/isaac/InkuluKap19}, in computing a Euclidean shortest path amid obstacles belonging to $\cal Q$, it suffices to consider only admissible cones to compute an $(1+\epsilon)$-approximate shortest path between any two points in $\cal{F(Q)}$ when $\epsilon \in (0, 0.6)$.
This optimization reduces the total number of cones introduced at the vertices of $\cal Q$.
This leads to introducing at most $O(\frac{1}{\epsilon})$ cones per obstacle.

\begin{figure}[htbp]
  \centering
  \includegraphics[width=0.5\textwidth]{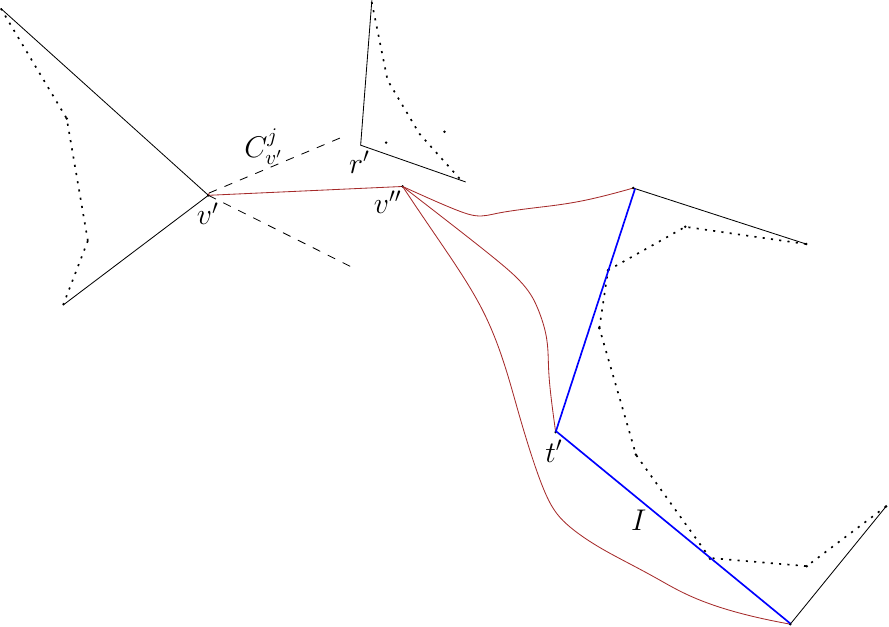}
  \caption{\footnotesize
   Illustrating an interval $I$ on the $bd({\cal Q})$ with respect to $v'$ in blue.
   The shortest paths from $v'$ to vertices belonging to $I$ are shown in brown.
   And the first edge $v'v''$ of the shortest path from $v'$ to every vertex of $I$ lies in an admissible cone $C_{v'}^j$.
   The closest visible vertex to $v'$ in cone $C_{v'}^j$ is $r'$, and hence we store $r'$ as the successor vertex along an approximate shortest path from $v'$ to any vertex of $I$.
   That is, if the destination $t'$ of $\cal Q$ belongs to $I$, then the line segment $v'r'$ belongs to the approximate shortest path being computed.
   The dotted black line segments are boundaries of convex polygonal obstacles in $\cal P$.
   \normalsize}
  \label{fig:interval}
\end{figure}

Next, we detail the satellite data that we store per vertex $v'$ of $\cal Q$ in a van Emde Boas tree $D_{v'}$.
At any vertex $v'$ of $\cal Q$, these additional data structures help in efficiently finding a vertex $v''$ of $\cal Q$ that occurs next to current vertex along an approximate shortest path from $v'$ to destination $t$ in $\cal Q$ such that the line segment $v'v''$ is in the $\cal{F(Q)}$. 
In computing an approximate shortest path query between any two query points $s$ and $t$ in $\cal{F(Q)}$, this algorithm outputs edges of that path by querying $D_{v'}$ associated with every vertex $v'$ of $\cal Q$ that occurs along that path.
For any vertex $v'$ of $\cal Q$, let $C_{v'}^j$ be an admissible cone with its apex at $v'$.
Let $S$ be the set of all vertices $w$ of $Q_i$ for which the first edge of the shortest path from $v'$ to $w$ lies in cone $C_{v'}^j$.
Due to the non-crossing property of the shortest paths, vertices in set $S$ are contiguous along the $bd(Q_i)$.
We call the maximal section of the $bd(Q_i)$ defined with vertices of $S$ an {\it interval} of $bd(Q_i)$ with respect to $v'$.
We note that all the intervals of $bd(Q_i)$ with respect to $v'$ together decompose the $bd(Q_t)$.
That is, each vertex of $bd(Q_i)$ belongs to an interval. 
For each interval $I$ defined with respect to any vertex $v'$ of $\cal Q$, if the shortest path tree edge corresponding to $I$ is incident to $v'$ and that edge belongs to admissible cone $C_{v'}^j$, then we store in the data structure associated to $v'$ the closest visible vertex $r'$ of $\cal Q$ in $C_{v'}^j$ together with $I$.
Refer to Fig.~\ref{fig:interval}.
Hence, the space required to store all the intervals defined with respect to all the vertices of $\cal Q$ together is $O(\frac{h}{\sqrt{\epsilon}} (h+\frac{1}{\epsilon}))$ and the number of admissible cones at all the vertices of $\cal Q$ together is $O(\frac{h}{\epsilon})$.
Therefore, the overall space is $O(\frac{h}{\sqrt{\epsilon}} (h+\frac{1}{\epsilon}))$. 
Following Banyassady~\etal~\cite{journals/cgta/BanyaCKMR20}, we partition the $bd({\cal Q})$ into intervals with respect to every vertex of $\cal Q$.
For every vertex $v'$ of $\cal Q$, the vertices of $\cal Q$ are partitioned into intervals with respect to $v'$. 
To compute all such intervals with respect to all the vertices of $\cal Q$, algorithm in \cite{journals/cgta/BanyaCKMR20} takes $O(\frac{h^2}{\epsilon} \lg \frac{h}{\sqrt{\epsilon}}+\frac{h}{\epsilon \sqrt{\epsilon}})$ time.
To compute the intervals corresponding to every vertex $v'$ of $\cal Q$, we construct the shortest path tree $T_{v'}$ in $\cal{F(Q)}$ at every vertex $v'$ of $\cal Q$ using the algorithm in Hershberger and Suri~\cite{journals/siamcomp/HershbergerS99}. 
This takes $O(\frac{h}{\sqrt{\epsilon}} \lg(\frac{h}{\sqrt{\epsilon}}))$.
For every obstacle $Q_i \in \cal Q$, and for every vertex $v'$ of $Q_i$, we compute a closest vertex $w'$ of $v'$ that is visible to $v'$ in every cone $C_{v'}^j \in  C(v')$.
For every interval $I$ defined with respect to a shortest path edge incident to $v'$ which is contained in $C_{v'}^j$, we associated $w'$ to $I$.
Given the shortest path tree rooted at $v'$, a nearest vertex in cone $C_{v'}^j$ can be computed using a plane sweep algorithm in $O(\frac{h}{\sqrt{\epsilon}} \lg(\frac{h}{\sqrt{\epsilon}}))$ time. 
Hence, all such plane sweeps for the shortest path tree rooted at each vertex of $\cal Q$ together take $O(\frac{h^2}{\epsilon} \lg(\frac{h}{\sqrt{\epsilon}}))$ time.
Since each of the vertices of $\cal P$ and each of the pockets is associated with a representative from $\cal Q$, we do not extend these interval structures on the $bd({\cal Q})$ to the $bd({\cal P})$.

We initialize an empty van Emde Boas tree $D_{v'}$ for every vertex $v'$ of $\cal Q$.
In the query phase, we use this vEB tree to efficiently find the next node along the shortest path from $v'$ to any vertex $t'$ of $\cal Q$.
Since $t'$ belongs to an interval of $v'$, the next vertex from $v'$ is fixed with respect to that interval and $v'$.
Consider an interval $\gamma$ located on the $bd(Q_\ell)$ with respect to $v' \in Q_i$ and cone $C_{v'}^j$ for some $j$.
Let $r'$ be the closest visible vertex of $\cal Q$ to $v'$ such that the line segment $v'r'$ is located in cone $C_{v'}^j$.
Let $u'$ and $w'$ be the endpoints of $\gamma$ such that $u'$ occurs before $w'$ in traversing the $bd(Q_\ell)$ in counterclockwise direction starting at $u'$, such that no vertex not in $\gamma$ is visited before visiting $w'$.
Instead of storing $\gamma$ and $r'$ in $D_{v'}$, we save the tuple consisting $w'$ and $r'$ in $D_{v'}$, wherein $w'$ is used as the key and $r'$ as its value.
Since the vertices of $\cal Q$ are partitioned into intervals, $w'$ essentially identifies the vertices belonging to $\gamma$.
For any vertex $t'$ to which a shortest path needs to be found from $v'$, instead of searching for $t'$ in $D_{v'}$, noting the identifier given to $t'$ lies between the identifiers of vertices $u'$ and $w'$, we search in $D_{v'}$ for $w'$.
This query returns $r'$.

\begin{lemma}
\label{lem:vEBanal}
The time to vEB trees associated to all the vertices of $\cal Q$ together takes $O(\frac{h}{\sqrt{\epsilon}} (h+\frac{1}{\epsilon}) \lg{\lg{\frac{h}{\sqrt{\epsilon}}}})$ time, 
and the space of all these vEB trees together is $O(\frac{h}{\sqrt{\epsilon}} (h+\frac{1}{\epsilon}))$.
\end{lemma}
\begin{proof}
The maximum value inserted into any of the $O(\frac{h}{\sqrt{\epsilon}})$ vEB trees is upper bounded by $O(\frac{h}{\sqrt{\epsilon}})$.
And, since there are in total $O(h+\frac{1}{\epsilon})$ number of intervals with respect to any vertex $v'$, the total time to insert all the key-value pairs into $D_{v'}$ is upper bounded by $O((h+\frac{1}{\epsilon}) \lg{\lg{\frac{h}{\sqrt{\epsilon}}}})$.
Since there are $O(\frac{h}{\sqrt{\epsilon}})$ vertices in $\cal Q$, the total time taken to insert all the key-value pairs into all the vEB trees together takes $O(\frac{h}{\sqrt{\epsilon}} (h+\frac{1}{\epsilon}) \lg{\lg{\frac{h}{\sqrt{\epsilon}}}})$ time.
And, the total space complexity of all vEB trees at all the vertices of $\cal Q$ together is $O(\frac{h}{\sqrt{\epsilon}} (h+\frac{1}{\epsilon}))$.
\end{proof}

Following Chen~\cite{conf/soda/Chen95}, we compute a set $\cal D$ of $O(\frac{1}{\epsilon})$ trapezoidal decompositions of $\cal{F(Q)}$.
That is, for every ray $\overrightarrow r$ bounding every cone in $\cal C$, at every vertex $v'$ of $\cal Q$, we introduce a maximal line segment in $\cal{F(Q)}$ parallel to $\overrightarrow r$ originating at $v'$.
All such rays together induce a trapezoidal decomposition of $\cal{F(Q)}$, and this subdivision is stored in ${\cal D}$.
Since $\cal Q$ has $O(\frac{h}{\sqrt{\epsilon}})$ vertices, each such decomposition is of size $O(\frac{h}{\sqrt{\epsilon}})$.
Since $|{\cal C}|$ is $\frac{1}{\epsilon}$ and since there is one trapezoidal decomposition corresponding to each ray bounding each of the cones in $\cal C$, the total space needed to store all the trapezoidal decompositions is $O(\frac{1}{\epsilon} \frac{h}{\sqrt{\epsilon}})$.
With a plane sweep for every one such ray, $\cal D$ can be computed in $O(\frac{1}{\epsilon} \frac{h}{\sqrt{\epsilon}} \lg(\frac{h}{\sqrt{\epsilon}}))$.
The decompositions stored in $\cal D$ together help in performing ray shooting from any point in $\cal{F(Q)}$ along any of the fixed directions defined by rays bounding cones in $\cal C$.
The query algorithm explains how this data structure helps determine whether any two given points $s$ and $t$ are mutually visible.

\begin{figure}[htbp]
  \centering
  \includegraphics[width=0.12\textwidth]{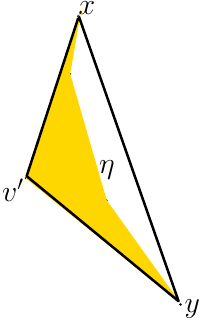}
  \caption{\footnotesize
  Illustrating the triangle $xv'y$ containing the pocket $\beta$ with representative $v'$, which is introduced into $\cal T$.
  Here $x, y$ are vertices of $\cal P$ and $v'$ is a vertex of $\cal Q$.
  Also showing the patch $\eta$ bounding $\beta$.
  \normalsize}
  \label{fig:augtriang}
\end{figure}

To determine whether each of the two input points $s$ and $t$ is located in $\cal{F(Q)}$ or $\cal{F(P)} \backslash \cal{F(Q)}$, our algorithm preprocesses $\cal{F(P)}$ to answer point location queries efficiently.
First, we triangulate $\cal{F(Q)}$ using the algorithm in \cite{journals/ijcga/Bar-YehudaC94}. 
This takes $O(\frac{h}{\sqrt{\epsilon}}+h \lg{h})$ time.
Let $\cal T$ be the set comprising all the triangles resulting from triangulating $\cal{F(Q)}$.
For every patch $\eta$, if $x, y$ are the vertices of $\cal P$ bounding $\eta$ and $v'$ is the representative vertex of $\eta$, then the triangle $xv'y$ is introduced into $\cal T$.
Essentially, this triangle contains the pocket corresponding to $\eta$.
Refer to Fig.~\ref{fig:augtriang}.
Since the number of patches on the $bd({\cal P})$ is $O(\frac{h}{\sqrt{\epsilon}})$, the total number of triangles in $\cal T$, excluding the triangles resulting from triangulating $\cal{F(Q)}$, is $O(\frac{h}{\sqrt{\epsilon}})$.
As a whole, the total number of triangles in $\cal T$ is $O(\frac{h}{\sqrt{\epsilon}})$.
Noting triangles in $\cal T$ are pairwise disjoint and together form a connected region, we compute Kirkpatrick's triangulation refinement~\cite{journals/siamcomp/Kirkpatrick83} based point location data structure for the triangles in $\cal T$. 
This data structure is of size $O(\frac{h}{\sqrt{\epsilon}})$ and is computed in $O(\frac{h}{\sqrt{\epsilon}} \lg{\frac{h}{\sqrt{\epsilon}}})$ time.
And, with this data structure, the point location query can be answered in $O(\lg{\frac{h}{\sqrt{\epsilon}}})$ time.
With each triangle $\Delta$ in the point location data structure, we store a flag indicating whether $\Delta$ lies in $\cal{F(Q)}$ or it lies in $Q_i \setminus P_i$ for some $i$. 
In the latter case, we also store the representative vertex of the pocket contained in $\Delta$.

As a whole, preprocessing involved computing the following: 
computing a sketch $\cal Q$ of $\cal P$,
partitioning the $bd({\cal P})$ into patches,
associating a representative vertex from $\cal Q$ to each patch,
partitioning $\cal{F(P)} \backslash \cal{F(Q)}$ into pockets,
computing a CVD in $\cal{F(Q)}$ per cone in $\cal C$,
partitioning the $bd({\cal Q})$ into intervals with respect to each vertex of $\cal Q$,
associating a vertex per interval while considering the admissible cones at the vertices of $\cal Q$,
initializing and storing tuples in vEB trees,
a set of trapezoidal decompositions, and
Kirkpatrick's point location data structure.
The following lemma upper bounds the cost of computing all these structures and the overall size of the data structures computed in the preprocessing phase.

\begin{lemma}
\label{lem:preproc}
The preprocessing algorithm computes data structures of size $O(n+\frac{h}{\sqrt{\epsilon}} (h+\frac{1}{\epsilon}))$ in $O(n+\frac{h}{\epsilon}(h+\frac{1}{\sqrt{\epsilon}})\lg(\frac{h}{\sqrt{\epsilon}}))$ time.
\end{lemma}

\section{The Query Algorithm}
\label{sect:query}

First, we determine whether $s$ and $t$ are visible.
For a cone $C_s$ that contains $t$, using the CVD corresponding to $C$, we determine a point $q$ on the $bd({\cal Q})$ that is closest to $s$ in $C_s$.
If $|st| < |sq|$, then $t$ is visible to $s$.
Otherwise, using preprocessed data structures $\cal D$ for trapezoidal decompositions of $\cal{F(Q)}$, we shoot a ray from $s$ along the direction of each of the rays bounding $C_s$ and let $e$ be the edge hit by either of these rays; then $t$ is not visible from $s$ if $e$ intersects the interior of $pq$.
Searching for a point $q$ in the CVD corresponding to $C$ and searching in the trapezoidal decompositions resulting from the line segments parallel to rays bounding $C$ takes $O(\lg{\frac{h}{\sqrt{\epsilon}}})$ time.
If not, again $t$ is visible to $q$.
In the cases in which $t$ is determined to be visible from $s$, we output the line segment $st$ as the shortest path.

\begin{figure}[htbp]
    \centerline{
    \includegraphics[width=0.5\linewidth]{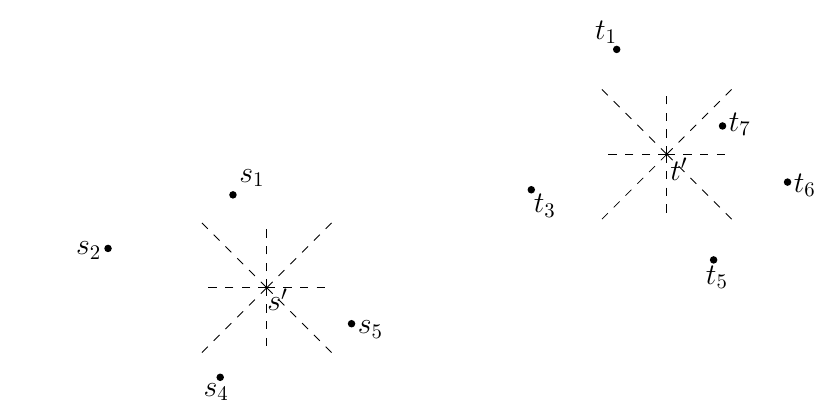}
    \hspace{0.25in}
    \includegraphics[width=0.45\linewidth]{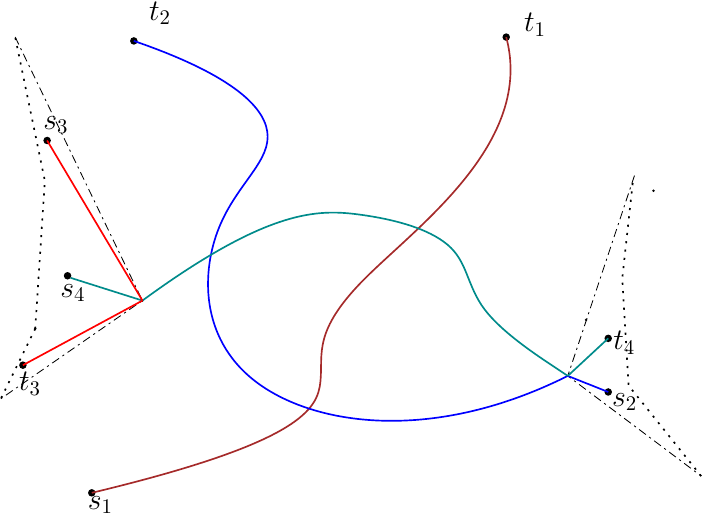}
    }
    \caption{\footnotesize
    Illustrating on the left a set $S'$ (resp., $T'$) of vertices of $\cal Q$ defined for $s'$ (resp., $t'$) located in $\cal{F(Q)}$.
    For every $j \in [1, |{\cal C}|]$, a point $s_j$ (resp., $t_j$) in $S'$ (resp., $T'$) corresponds to a closest vertex of $Q$ that is visible to $s'$ (resp., $t'$) in cone $C_{s'}^j$ (resp., $C_{t'}^j$).
    The right illustration shows cases being handled in the query algorithm: (i) both the query points $s_1, t_1 \in \cal{F(Q)}$, (ii) $s_2 \in {\cal F(P)} \backslash {\cal F(Q)}, t_2 \in \cal{F(Q)}$, (iii) both $s_3, t_3$ belong to a pocket, and (iv) both $s_4$ and $t_4$ belong to distinct pockets.
    The corresponding approximate shortest paths are also depicted in distinct colours.
        \normalsize}
    \label{fig:queryalgo}
\end{figure}

Using the point-location data structure, the query algorithm determines the triangle $\Delta_s$ (resp., $\Delta_t$) containing $s$ (resp., $t$).
The query algorithm then handles the following cases.
Refer to Fig.~\ref{fig:queryalgo}.

In Case~(i), both $s$ and $t$ lie in $\cal{F(Q)}$, 
For each cone $C \in \cal C$, using the CVD corresponding to $C$, we find the closest vertex $s'$ of $\cal Q$ in $C_s$ such that $s'$ is visible to $s$.
Analogously, for each cone $C \in \cal C$, we find the closest vertex $t'$ of $\cal Q$ in $C_t$ such that $t'$ is visible to $t$.
Let $S'$ (resp., $T'$) be the set comprising all such nearest visible vertices to $s$ (resp., $t$).
For every $s' \in S'$ and $t' \in T'$, with an iterative algorithm, we compute an approximate shortest path between $s'$ and $t'$: starting with $s'$, at every vertex $v'$ along the path being computed, we inquire $D_{v'}$ with destination vertex $t'$ of $\cal Q$, to find the next vertex.
And, output a path of minimum length among these paths.

In Case~(ii), $s \in \cal{F(P)} \backslash \cal{F(Q)}$ and $t \in \cal{F(Q)}$ (the other case in which $s \in \cal{F(Q)}$ and $t \in \cal{F(P)} \backslash \cal{F(Q)}$ is handled analogously), and
Since $s$ is located in a triangle $\Delta_s$ (introduced into the point location data structure $L$) corresponding to a pocket, from $\Delta_s$ in $L$, we extract the representative vertex $s'$ of $\Delta_s$.
For $t$, as in Case~(i), for each cone $C \in \cal C$, we find the closest vertex $t'$ of $\cal Q$ in $C_t$ such that $t'$ is visible to $t$.
Let $T'$ be the set comprising all such nearest visible vertices to $t$.
Like in Case~(i), for every $t' \in T'$, we compute an approximate shortest path between $s'$ and $t'$, and output the one among these that has the shortest length with the line segment $ss'$ concatenated at $s'$.

In Case~(iii), both $s$ and $t$ belong to the same pocket, say $\Delta$. 
First, we find the representative vertex $s'$ of $\Delta$.
Then we output the line segments $ss'$ and $s't$ as the approximate shortest path from $s$ to $t$.
Let $p, q,$ and $s'$ be the vertices of $\Delta$.
Then, the length of this two-segment path is equal to 
$|ss'| + |s't|
\le |ps'|+|qs'|
\le \frac{|pq|}{\sin(\tfrac{\pi}{2}- \sqrt{\frac{\epsilon}{2})}}
\le(1+\epsilon)|pq|
\le (1+\epsilon)d_{\cal P}(s,t)$.

In Case~(iv), both $s$ and $t$ belong to interiors of distinct pockets.
Here, we find the representative vertex $s'$ (resp., $t'$) of pocket in which $s$ (resp., $t$) lies from $\Delta_s$ (resp., $\Delta_t$).
And we compute an approximate shortest path $\pi_{s't'}$ from $s'$ to $t'$ in $\cal Q$, and output the concatenation of line segment $ss'$, $\pi_{s't'}$, and the line segment $t't$ as the approximate shortest path from $s$ to $t$ in $\cal{F(P)}$.
Let $s'=v_1', v_2', \ldots, v_{k-1}', v_k'=t'$ be the sub-path of the path output by the algorithm.
Note that all these vertices are vertices of $\cal Q$.
Let $\ell$ be the maximum length of any patch in $\cal P$.
Then, it is immediate to note that $\ell \le (\sqrt{2\epsilon}) (\max_{P_i \in \cal P} diameter(P_i))$, where $diameter(P_i) =$ $\max_{p,q \in P_i} |pq|$.
For any two points $p, q$ belonging to any pocket, from the above analysis, the distance $\lambda$ along the approximate shortest path between $p$ and $q$ output by this algorithm is upper bounded by $(1+\epsilon) \ell$.
Also, let $k$ be the number of line segments in the output path.
Then, the length of the path between $s$ and $t$ output by the algorithm is equal to, 
\begin{flalign*}
\hspace{0.7in}
&|ss'| + (\sum_{i=1}^{k-1}( |v_i' v_{i+1}'|)) + |t't| && \\
&\le  2\lambda+ (\sum_{i=1}^{k-1}( |v_i' v_{i+1}'|))  &&\\
&\hspace{0.1in} \text{[since $d'_{\cal P}(s,v_1') \le \lambda$, $d'_{\cal P}(v_k',t) \le \lambda $]} && \\[6pt]
&\le  2\lambda + (1 + \epsilon)  d_{\cal{Q}}(v_1', v_k')   && \\
&\hspace{0.1in} \text{[from Clarkson~\cite{conf/stoc/Clarkson87}, $( \sum_{i=1}^{k-1}( |v_i' v_{i+1}'|) \le (1 + \epsilon)  d_{\cal{Q}}(v_1', v_k')$]} &&\\
&\le 2\lambda + (1 + \epsilon)^2   d_{\cal P}(v_1', v_k') && \\
&\hspace{0.1in} \text{[from Agarwal~\etal~\cite{conf/soda/AgarwalSY09}, $d_{\cal{Q}}(v_1', v_k') \le (1 + \epsilon)  d_{\cal P}(v_1', v_k')$]} && \\[6pt]
&\le  2\lambda + (1 + \epsilon)^2   d_{\cal P}(s, t) + (1 + \epsilon)^2  ( d(s,v_1') + d(v_k',t) ) && \\
&\hspace{0.1in} \text{[by the triangle inequality]} && \\[2pt]
&\le 2\lambda + (1 + \epsilon)^2 d_{\cal P}(s, t) + (1 + \epsilon)^2  ( 2\lambda )  && \\
&\hspace{0.1in} \text{[since $d(s,v_1') <\lambda$  and $d(v_k',t) <\lambda$]} && \\[2pt]
&\le (1 + \epsilon)^2   d_{\cal P}(s, t) + (2+ 3\epsilon)  (2\lambda)   && \\
	&\le (1 + \epsilon)^2   d_{\cal P}(s, t) + 13\ell && \\
&\hspace{0.1in} \text{[since $\epsilon \in (0, 0.6)$]} && \\[2pt]
&\le (1 + 3\epsilon)   d_{\cal P}(s, t) + 13\ell.
\end{flalign*}
With an analogous analysis, it can be shown that the stretch obtained in this case upper bounds the stretch resultant from either Case~(i) or Case~(ii).

\begin{lemma}
\label{lem:stretch}
The approximate shortest path computed has $(1+\epsilon)$ multiplicative stretch and $13 \ell$ additive stretch.
\end{lemma}

\begin{lemma}
\label{lem:aspinG}
Computing an approximate shortest path in $\cal Q$ between any two vertices $s', t'$ of $\cal Q$ takes $O(k (\lg{\lg(\frac{h}{\sqrt{\epsilon}})}))$ time. 
Here, $k$ is the number of line segments in the path output, and $k$ is upper bounded by $O(\frac{h}{\sqrt{\epsilon}})$.
\end{lemma}
\begin{proof}
Let $v'$ be any intermediate node on the approximate shortest path between $s'$ and $t'$.
Let $I$ be an interval defined with respect to $v'$ and $t$ belongs to $I$. 
Also, let $u'$ and $w'$ be the first and last vertices of $I$ such that $u'$ occurs before $w'$ while traversing $I$ in counterclockwise direction.
We search for the successor $w'$ of $t'$ at $D_{v'}$ with key $t'$.
As mentioned, since the number of vertices of $\cal Q$ is upper bounded by $\frac{h}{\sqrt{\epsilon}}$, the universe size of any vEB structure stored at any vertex of $\cal Q$ is $O(\frac{h}{\sqrt{\epsilon}})$.
Hence, to find the key $w'$ that is the successor of $t'$ in vEB $D_{v'}$ takes $O(\lg \lg (\frac{h}{\sqrt{\epsilon}}))$ time.
The value associated with $w'$, which is the identifier of the next vertex on the approximate shortest path from $v'$ to $t'$, is extracted from the key-value pair in $O(1)$ time.
If there are $k$ vertices that belong to any approximate shortest path, then the total time taken is as stated. 
\end{proof}

\begin{lemma}
The algorithm takes $O(\frac{1}{\sqrt{\epsilon}}(\lg{\frac{h}{\sqrt{\epsilon}}}) + \frac{h}{\epsilon^{2.5}} (\lg \lg (\frac{h}{\sqrt{\epsilon}})))$ time to compute an approximate shortest path between any two query points $s$ and $t$ in $\cal{F(P)}$. 
\end{lemma}
\begin{proof}
Using the point location data structure, locating both $s$ and $t$ together takes $O(\lg{\frac{h}{\sqrt{\epsilon}}})$ time.
Further, determining whether $s$ belongs to a pocket takes $O(1)$ time.
If it is, retrieving the representative vertex of that pocket takes $O(1)$ time.
Considering all the cases, at $s$, searching in one CVD corresponding to each cone in $\cal C$ to find a nearest visible point $s'$ to $s$, which is a vertex of $\cal Q$ takes $O(\lg{\frac{h}{\sqrt{\epsilon}}})$.
Hence, the set $S'$ comprising all such $s'$ points together is of size $O(|{\cal C}|)$, which is $O(\frac{1}{\epsilon})$.
The same is the size of $T'$.
From Lemma~\ref{lem:aspinG}, computing a shortest path between any $s'$ and $t'$ takes $O(\frac{h}{\sqrt{\epsilon}}(\lg{\lg{(\frac{h}{\sqrt{\epsilon}})}}))$ time.
And, since $|S' \times T'|$ is $O(\frac{1}{\epsilon^2})$, there are $O(\frac{1}{\epsilon^2})$ shortest path computations in $\cal Q$.
\end{proof}

\setcounter{theorem}{0}

\begin{theorem}
Given a convex polygonal domain $\cal P$ comprising $h$ convex polygonal obstacles defined with $n$ vertices and a real number $\epsilon \in (0, 0.6)$, in $O(n+\frac{h}{\epsilon}(h+\frac{1}{\sqrt{\epsilon}})\lg(\frac{h}{\sqrt{\epsilon}}))$ time, the preprocessing algorithm computes data structures of size $O(n+\frac{h}{\sqrt{\epsilon}} (h+\frac{1}{\epsilon}))$, so that to output in time $O(\frac{1}{\sqrt{\epsilon}}(\lg{\frac{h}{\sqrt{\epsilon}}})+\frac{h}{\epsilon^{2.5}}(\lg{\lg(\frac{h}{\sqrt{\epsilon}})}))$ an approximate shortest path between any two query points in $\cal{F(P)}$ wherein that path has multiplicative stretch $1+\epsilon$ and an additive stretch $13 \ell$.
Here, $\ell$ is upper bounded by $(\sqrt{2\epsilon})(\max_{P_i \in {\cal P}} \max_{p, q \in P_i} |pq|)$.
\end{theorem}

\section{Conclusions}
\label{sect:conclu}

Given a polygonal domain $\cal P$ with $h$ convex obstacles defined by $n$ vertices and a parameter $\epsilon$ in $(0, 0.6)$, the approximation scheme presented herewith preprocesses $\cal P$ to facilitate answering two-point approximate shortest paths.
The space of data structures, the preprocessing time to compute them, and the query time are improved for convex polygonal obstacles when $h < n$, which is typically the case. 
The multiplicative stretch of the path output is $(1+\epsilon)$, but it has an additive stretch as well.
Future work could consider removing the additive stretch factor and extending some of these ideas to simple polygonal domains.

\subsection*{Acknowledgements}

This research of R. Inkulu is supported in part by the National Board for Higher Mathematics (NBHM) grant 2011/33/2023NBHM-R\&D-II/16198.

\bibliographystyle{plain}
%\bibliography{../ajar/results/bibs/geomgraphs,../ajar/results/bibs/misc,../ajar/results/bibs/shortestpaths,../ajar/results/bibs/visibility,../ajar/results/bibs/routing}

\end{document}